\documentclass[aps,prb,twocolumn,preprintnumbers,amsmath,amssymb,nofootinbib,floatfix]{revtex4}
\pdfoutput=1
\usepackage{graphicx}		
\usepackage{dcolumn}		
\usepackage{bm}			

\begin{document}

\title{Electronic and lattice dynamical properties of the iron-based superconductors LiFeAs and NaFeAs}
\author{R. A. Jishi}
\author{H. M. Alyahyaei}
\affiliation{ Department of Physics, California State University, Los Angeles, California 90032 }

\date{\today}

\begin{abstract}
The electronic structure and lattice vibrational frequencies of the newly discovered 
superconductors, LiFeAs and NaFeAs, are calculated within density functional theory. 
We show that in the vicinity of the Fermi energy, the density of states is dominated 
by contributions from Fe 3d states. We also calculate the electron-phonon coupling 
strength and show that it is too weak to account for the observed values of the 
superconducting transition temperatures. This seems to indicate that the iron-based 
superconductors are not of the conventional type.

\end{abstract}

\maketitle

\section{\label{sec:introduction}Introduction}
A new class of layered, high-T$_{c}$ superconductors has been recently
discovered. Kamihara et al.~\cite{Kamihara_1:2006,Kamihara_2:2008} reported a superconducting
transition temperature T$_{c}$=26 K in fluorine-doped LaOFeAs. Shortly
afterwards, it was found that under pressure T$_{c}$ increased to 43 K.~\cite{Takahashi:2008} Replacement of lanthanum with other
rare earth metals gave a series of superconducting compounds ReO$_{1-x}$F$_{x}$FeAs,
where Re = Ce, Pr, Nd, Sm, or Gd, with transition temperatures close to or
exceeding 50 K.~\cite{Chen_1:2008,Ren_1:2008,Ren_2:2008,Chen_2:2008,Ren_3:2008,Cheng:2008} Oxygen deficient samples were also
synthesized and found to superconduct at 55 K.~\cite{Yang:2008,Wu_1:2008,Ren_4:2008} Hole doping,
through the partial substitution of La with Sr, or Gd with Th, was also
found to yield superconducting compounds.~\cite{Wen:2008,Wang_1:2008} Using high pressure
techniques, it was possible to increase the concentration of the
F-dopant~\cite{Lu:2008} and to synthesize superconducting compounds where La is
replaced by the late rare earth elements Tb and Dy.~\cite{Bos:2008,Li:2008}
The parent compound ReOFeAs is a layered
compound consisting of a stack of alternating ReO and FeAs layers. Each ReO
layer consists of an O-sheet surrounded by two Re sheets. Similarly, each FeAs
layer consists of an Fe-sheet surrounded by two As sheets such that each Fe
atom is tetrahedrally coordinated to four As atoms. Neutron diffraction 
measurements~\cite{Cruz:2008,Zhao_1:2008,Zhao_2:2008,Qiu:2008}establish that the Fe magnetic moments 
adopt a collinear antiferromagnetic (c-AFM) order whereby
ferromagnetic chains are coupled antiferromagnetically along the 
direction orthogonal to the chains.

Superconductivity was also discovered in a second class of compounds
containing FeAs layers, namely AFe$_{2}$As$_{2}$, where A is an alkaline
earth metal. Hole doping, by partial replacement of A with alkali metals, 
results in superconducting compounds with T$_{c}$ reaching 38 K in 
BaFe$_{2}$As$_{2}$ and SrFe$_{2}$As$_{2}$.\cite{Rotter:2008,Chen_3:2008,Sasmal:2008,Wu:2008,Ni:2008} 
Partial substitution of Fe with Co was also shown to give a 
superconducting compound with T$_{c}$=22 K.~\cite{Sefat:2008}
Similarly to the first class, in the parent compounds the Fe magnetic 
moments in this second class have a collinear AFM order with a 
spin-stripes pattern.~\cite{Huang:2008,Kitagawa:2008,Su:2008} 
In both classes, the Fe magnetic moments in the parent compounds exhibit 
magnetic order, at low temperature, which disappears upon doping, making way 
for the emergence of superconductivity. This leads to the reasonable 
belief that strong electronic correlations are important in these systems, 
and that superconductivity in these compounds is somehow connected to 
magnetic fluctuations.~\cite{Haule:2008,Singh_1:2008,Ma:2008,Dong:2008,Cao:2008,Yildirim:2008,Weng:2008,Ma_1:2008,Yin:2008,Ishibashi:2008,Si:2008,Fang:2008,Xu:2008} Indeed, the
electron-phonon coupling in LaOFeAs was estimated to be too small~\cite{Boeri:2008} to
give rise to superconductivity within the conventional BCS formulation.~\cite{Bardeen:1957}

Recently, a third class of iron-based superconductors was discovered. LiFeAs
and NaFeAs were found to superconduct below 18 K and 9 K, respectively.~\cite{Pitcher:2008,Tapp:2008,Wang:2008,Parker:2008} It turns out 
that in these two compounds, no magnetic order is detected
at all temperatures. In some sense, these two compounds are important with
regards to understanding the mechanism of superconductivity in iron-based
superconductors. the absence of spin density wave (SDW) transition, on the
one hand, and the relatively low T$_{c}$ in comparison with the first two classes
of iron-based superconductors, on the other hand, make these two compounds
possible candidates for being conventional BCS superconductors.

Band structure calculations, using local density approximation (LDA) within
density functional theory (DFT), were recently reported for LiFeAS.~\cite{Nekrasov:2008,Singh:2008}
It was found that LiFeAs is semi-metallic, and that the density of states (DOS) near 
the Fermi level is dominated by the Fe 3d states. Thus, the electronic structure of
stoichiometric LiFeAs is similar to that of the parent compounds of the first class,
with a hole cylinder at the Brillouin zone (BZ) center, electron cylinders at the BZ 
corners, and an electronic DOS that decreases strongly with increasing energy in the
vicinity of the Fermi energy.

In this work we report DFT calculations of the
electronic and lattice properties of LiFeAs and NaFeAs. In particular, we
calculate the electron-phonon coupling strength and show that it is too
weak to account for the superconducting transition temperatures observed in
these compounds. Our calculations, together with previous calculations~\cite{Boeri:2008}
of the electron-phonon coupling strength in LaOFeAs, seem to indicate that
iron-based superconductors are not of the conventional type.

\section{\label{sec:method}Method}

The electronic structure calculations are carried out using the 
all-electron full-potential
linear augmented plane wave (FP-LAPW) method as implemented in WIEN2K code.~\cite{Blaha:2001}
The exchange-correlation potential was calculated using
the generalized gradient approximation (GGA) as proposed by Pedrew,
Burke, and Ernzerhof (PBE).~\cite{Perdew:1996} The radii of the muffin-tin 
spheres for the various atoms were chosen so that the neighboring
spheres almost touch each other. We set the parameter R$_{MT}$K$_{max}$=7,
where R$_{MT}$ is the smallest muffin-tin radius and K$_{max}$ is a
cutoff wave vector. The valence wave functions inside the muffin-tin
spheres are expanded in terms of spherical harmonics up to $l_{max}=10$,
and in terms of plane waves with a wave vector cutoff K$_{max}$ in the
interstitial region. The charge density is Fourier expanded up to
G$_{max}$=13$a_{0}$$^{-1}$, where $a_{0}$ is the Bohr radius. 
Convergence of the self-consistent field calculations is attained with
a total energy convergence tolerance of 0.01 mRy.  

The calculation of the frequencies of the vibrational modes and 
the electron-phonon coupling parameter
was performed using ultrasoft pseudopotentials and an expansion of
the wave function of the valence electrons in terms of plane waves,
with an energy cutoff of 30 Rydbergs.~\cite{Giannozzi:2006}
In both the electronic and lattice calculations, the experimental 
values of the low-temperature lattice constants and atomic
positions~\cite{Pitcher:2008,Parker:2008} are used. 
For both compounds, the
crystal is tetragonal with space group P4/nmm. In LiFeAs, the lattice constants
are $a$ = 3.76982 $\AA$, $c$ = 6.30693 $\AA$, whereas in NaFeAs, 
$a$ = 3.94729$\AA$, and $c$ = 6.99112 $\AA$.

\section{\label{sec:results_and_discussion}Results and Discussion}
\begin{figure*}
   \includegraphics[width=\textwidth]{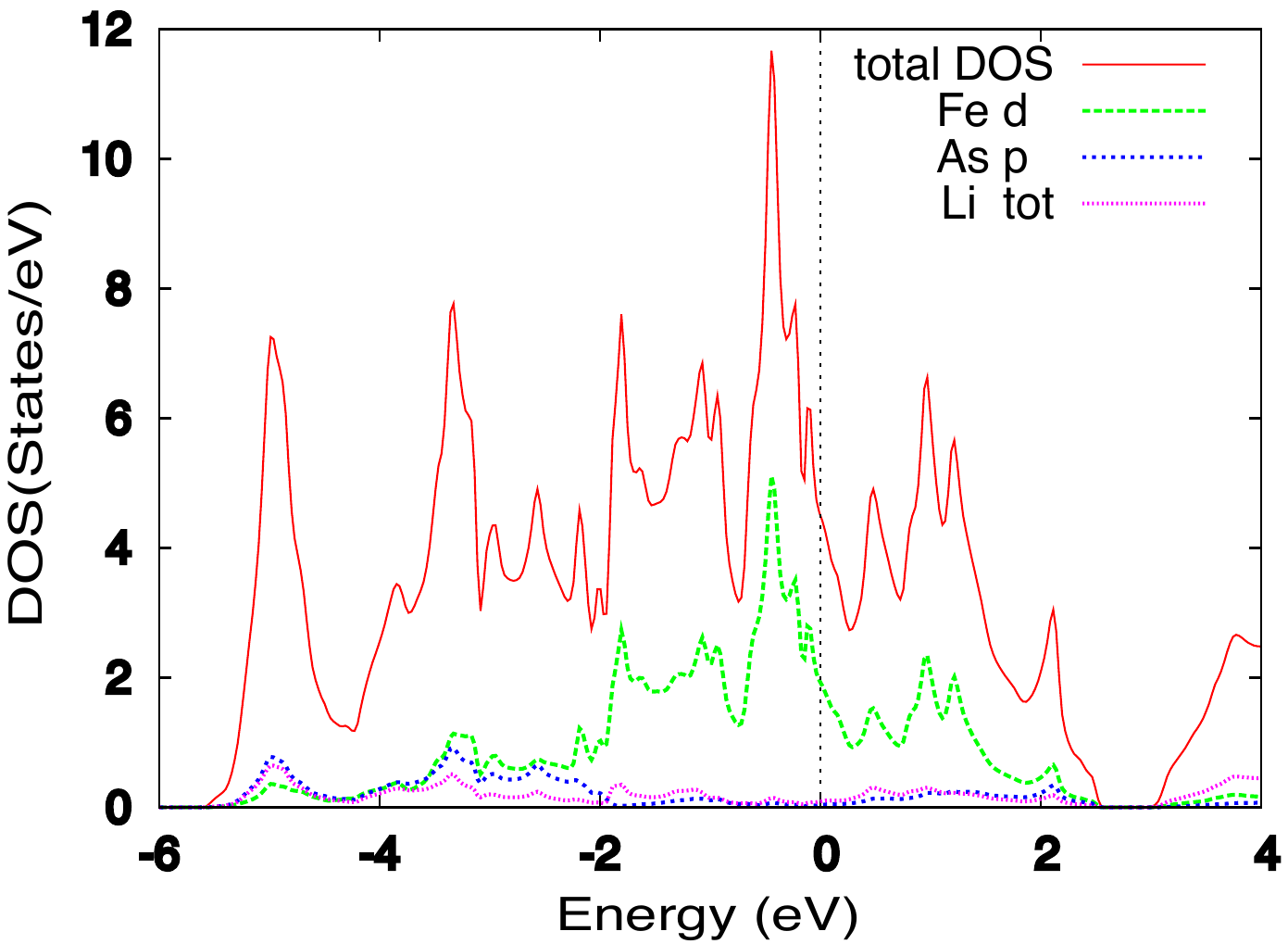}
   \caption{\label{fig:figure1}(Color online) Density of states (DOS) in LiFeAs. Both the total 
and atomic DOS are shown. The Fermi energy 
is the zero energy. Near the Fermi energy, the DOS is dominated by the Fe 3d states.}
\end{figure*}

\begin{figure*}
   \includegraphics[width=\textwidth]{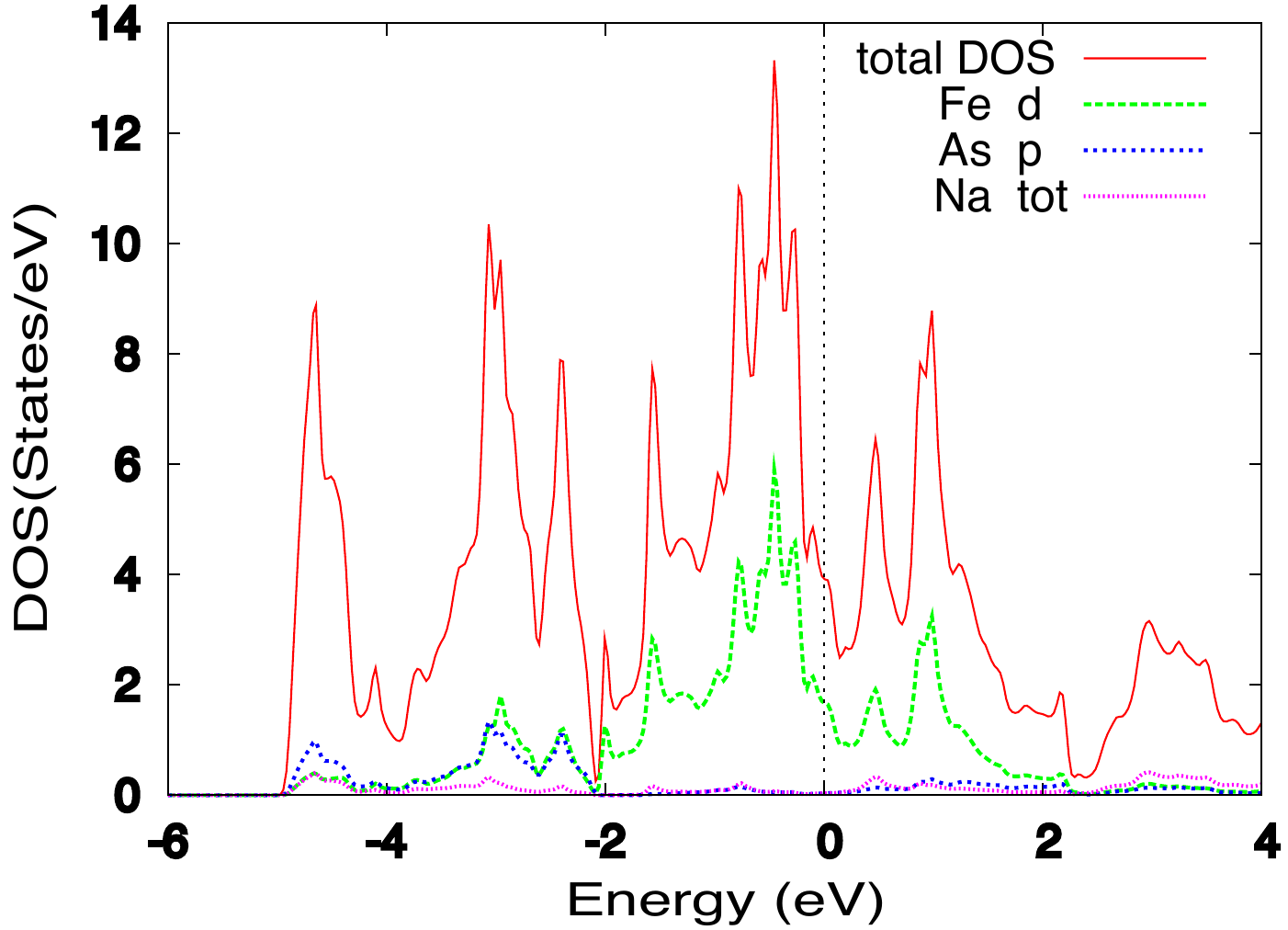}
   \caption{\label{fig:figure2}(Color online) Density of states (DOS) in NaFeAs. Both the total
and atomic DOS are shown. The Fermi energy 
is the zero energy. Near the Fermi energy, the DOS is dominated by the Fe 3d states. }
\end{figure*}

Our results for the electronic structure calculations for LiFeAs and NaFeAs 
are summarized in figures~\ref{fig:figure1} and~\ref{fig:figure2}, respectively, 
where the electronic density
of states (DOS) is displayed. For LiFeAs, our calculated DOS is similar to
that reported earlier.~\cite{Nekrasov:2008,Singh:2008} In both LiFeAs and NaFeAs,
the DOS plots show some generic features that are common to the parent
compounds of the iron-based superconductors:
A DOS that is dominated by the Fe 3d states in the vicinity of the 
Fermi energy, with only a small contribution from the As and alkali 
metal states, and that is strongly decreasing with energy near the 
Fermi energy. It should be noted that in a unit cell of LiFeAs, for 
example, there are two Li, two Fe, and two As atoms. Thus to get the
total atomic DOS, the values of the atomic DOS shown in~\ref{fig:figure1} should
be multiplied by 2. The total DOS is the sum of the total atomic DOS and
the DOS in the interstitial region.

Since the DOS at the Fermi energy, N(E$_{F}$), is $\sim$4 states/eV in both LiFeAs
and NaFeAs, which is not very small, and because of the relatively lower T$_{c}$
compared with the other iron-based superconductors, one may wonder whether
electron-phonon coupling may lie behind the mechanism for superconductivity
in LiFeAs and NaFeAs. This notion may be given more credence by the
observation that a sodium atom is about three times more massive than a
lithium atom, so that if the attractive electron-electron interaction is
mediated by the alkali-metal atomic vibrations, then this difference in
the mass could explain the difference in the values of T$_{c}$ between the two
compounds via the well-known isotope effect.

\begin{figure*}
   \includegraphics[width=\textwidth]{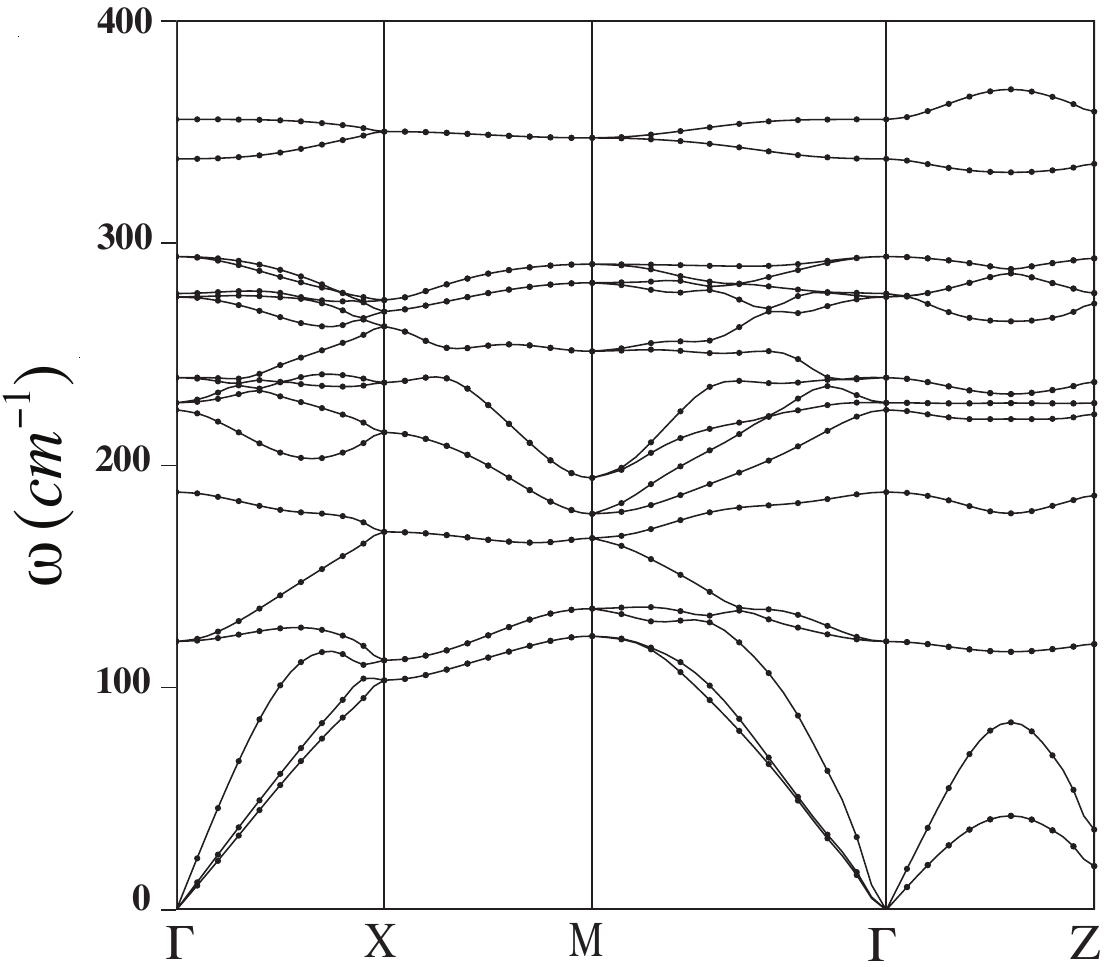}
   \caption{\label{fig:figure3} Phonon dispersion curves in LiFeAs, plotted along high symmetry directions of the Brillouin zone.}
\end{figure*}

\begin{figure*}
   \includegraphics[width=\textwidth]{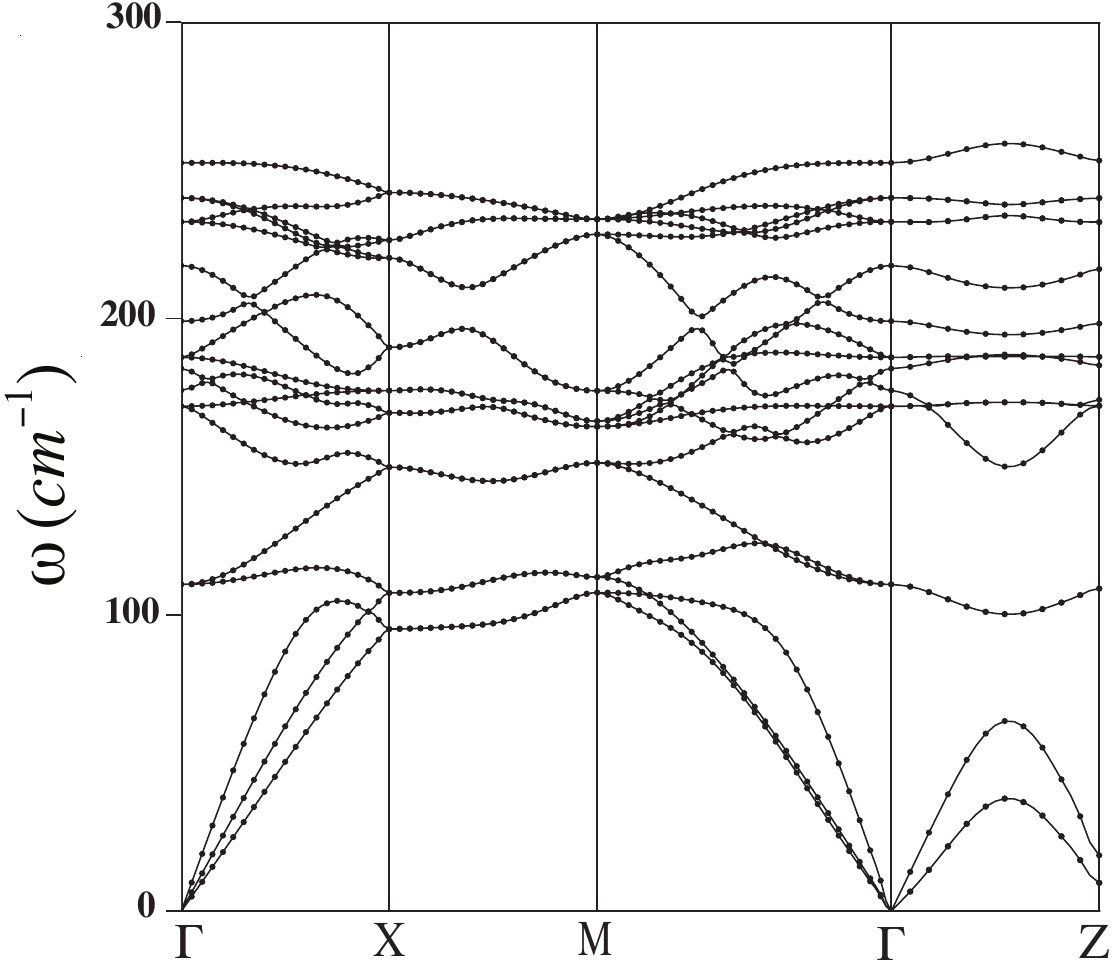}
   \caption{\label{fig:figure4} Phonon dispersion curves in NaFeAs, plotted along high symmetry directions of the Brillouin zone.}
\end{figure*}

To test this idea, we carried out a calculation of the phonon dispersion
curves and the electron-phonon coupling strength in these compounds. 
Since the crystallographic point group in LiFeAs and NaFeAs is D$_{4h}$,
the vibrational modes at $\Gamma$, the BZ center, are decomposed 
according to the following irreducible representations 

\[
\Gamma_{phonon} = 2A_{1g} + B_{1g} + 3E_{g} + 3A_{2u} + 3E_{u} .
\]
\begin{table*}
        \caption{\label{tab:Phonon}The calculated frequencies, in cm$^{-1}$, of the Raman- and infrared (IR)-active modes in LiFeAs
and NaFeAs. The modes are classified by the irreducible representations (irreps) according to which they transform.}
        \begin{ruledtabular}
        \begin{tabular}{l r|c c c c c c}
\multicolumn{3}{c}{\ } & \multicolumn{2}{c}{$\omega$(irrep)/cm$^{-1}$}&\multicolumn{2}{c}{\ }   \\\hline
          &Raman&121(E$_{g}$)&188(A$_{1g}$)&225(B$_{1g}$)&240(E$_{g}$)&294(E$_{g}$)&356(A$_{1g}$) \\
   LiFeAs &     &            &             &             &             \\
          &IR   &228(E$_{u}$)&276(E$_{u}$) &277(A$_{2u}$)&338(A$_{2u}$)\\ \hline
          &Raman&110(E$_{g}$)&176(A$_{1g}$)&187(E$_{g}$) &199(A$_{1g}$)&218(B$_{1g}$)&241(E$_{g}$) \\
   NaFeAs &     &      &     &      &       \\
          &IR   &170(E$_{u}$)&183(A$_{2u}$)&233(E$_{u}$)&253(A$_{2u}$)\\
          \end{tabular}
        \end{ruledtabular}
\end{table*}
The acoustic modes, with vanishing frequency at $\Gamma$, the BZ center, transform 
according to the A$_{2u}$ and E$_{u}$ irreducible representations. 
Excluding the acoustic modes, we are left with 15 modes with nonzero
frequencies; among these, the symmetric ones are Raman-active, while
the antisymmetric modes are infrared-active. 
The calculated frequencies of the Raman- and infrared-active modes at 
the $\Gamma$ point of the BZ are given in Table~\ref{tab:Phonon}. 
The phonon dispersion curves in LiFeAs, plotted along
high symmetry directions in the BZ, are shown in Fig.~\ref{fig:figure3}, 
and the corresponding curves in NaFeAs are given in Fig.~\ref{fig:figure4}.
Our results for the phonon frequencies at the BZ center may be checked
by Raman scattering and infrared absorption experiments, while the phonon
dispersion curves may be checked by neutron scattering measurements.

We calculated the electron-phonon coupling parameter $\lambda$ and found it to
be 0.29 and 0.27 for LiFeAs and NaFeAs, respectively. For conventional
superconductors, where the attractive electron-electron interaction is
mediated by phonons, the transition temperature is given by the Allen 
and Dynes~\cite{Allen:1975} 
modified McMillan's formula~\cite{McMillan:1968}

\[
T_{c} = \frac{\omega_{log}}{1.2}exp[\frac{-1.04(1+\lambda)}{\lambda-\mu^{*}(1+0.62\lambda)}]
\]

where $\omega_{log}$ is the logarithmic average phonon frequency, 
expressed in degrees Kelvin,
and $\mu^{*}$ is the Coulomb pseudopotential parameter, usually taken 
to be $\sim$0.13. Since $\omega_{log}$ in LiFeAs and NaFeAs is of the
order of 100 K, the resulting value of T$_{c}$ is much less than 1 K. We conclude
that the electron-phonon coupling is too weak to account for superconductivity
in this class of iron-based superconductors.

There are some puzzling questions that beset this third class of iron-based
superconductors. In the parent compounds of the first two classes of
iron-based superconductors, magnetic order is established at low temperatures,
where the Fe magnetic moments adopt a collinear antiferromagnetic (c-AFM)
 order; this is
unequivocally confirmed by both neutron diffraction measurements~\cite{Cruz:2008,Zhao_1:2008,Zhao_2:2008,Qiu:2008} and
DFT calculations.~\cite{Ishibashi:2008,Dong:2008,Ma:2008,Haule:2008,Cao:2008,Yildirim:2008,Weng:2008,Ma_1:2008,Yin:2008} 
In the first class, it is only upon electron doping through the replacement of a small percentage
of oxygen atoms with fluorine atoms, or the removal of a small percentage
of oxygen atoms, that the magnetic order is suppressed and superconductivity
emerges. In the second class of iron-based superconductors, magnetic order
is suppressed by hole doping through the replacement of some alkaline
earth atoms with alkali atoms. To better understand the situation in 
the third class of iron-based superconductors, we carried out spin-polarized
DFT calculations on stoichiometric LiFeAs and NaFeAs, in addition to 
the calculations reported above for the nonmagnetic
phases of these compounds. We considered, 
within GGA and GGA+U, 
various spin arrangements on the Fe sites. Similarly to
the case of the first two classes, we find that the c-AFM phase,
with a spin-stripes pattern, is indeed the lowest energy phase. Within
GGA, the energy of the c-AFM phase in LiFeAs is lower than the AFM
phase by 0.081 eV per Fe atom (eV/Fe), lower than the ferromagnetic (FM) 
phase by 0.085 eV/Fe, and lower than the nonmagnetic phase 
by 0.123 eV/Fe. For NaFeAs, the energy of the c-AFM
phase is lower than the AFM phase by 0.048 eV/Fe, lower than the FM phase
by 0.205 eV/Fe, and lower than the nonmagnetic phase by 0.182 eV/Fe. 
The differences are even greater within GGA+U, where onsite Coulomb 
repulsion is taken into account. Thus, according to our DFT calculations, 
stoichiometric LiFeAs and NaFeAs should be similar to the parent compounds of the
first two classes, and they should not superconduct; instead, at low 
temperature, the stoichiometric compounds should display magnetic order.
Deviations from stoichiometry, on the other hand, may suppress the 
magnetic order, making way for superconductivity, just like  
doping does in the first two classes of iron-based compounds. In the 
case of LiFeAs, it is indeed the case that the synthesized 
superconducting compounds were not stoichiometric, the chemical formula 
being Li$_{1-x}$FeAs.~\cite{Wang:2008} The situation is less clear in the case of 
NaFeAs,~\cite{Parker:2008} but we believe, on the basis of our spin-polarized 
calculations, and the absence of any detectable magnetic order at low temperatures, that 
the synthesized NaFeAs samples must also be nonstoichiometric, and that 
further studies on sample characterization are necessary in this regards.

\section{\label{sec:conclusion}Conclusions}

In conclusion, we have presented the results of electronic structure
calculations on LiFeAs and NaFeAs, members of a new class of superconducting
compounds. In similarity to other iron-based superconductors, the density of
states in the vicinity of the Fermi energy is found to be dominated by
contributions from the Fe 3d states. We have also calculated the Raman
and infrared phonon frequencies at the Brillouin zone center, as well 
the phonon dispersion curves along high symmetry directions in the
Brillouin zone. We have evaluated the electron-phonon coupling parameter
in LiFeAs and NaFeAs, and found its value to be too small to account
for the observed superconducting transition temperatures in these
compounds. Our results, taken together with previous estimates of the
electron-phonon coupling strength in LaOFeAs, seem to suggest clearly
that iron-based superconductors are not of the conventional type, where
the attractive electron-electron interaction is mediated by phonons. 


\end{document}